\documentclass[aps,prb,twocolumn,superscriptaddress,amsmath,amssymb,showpacs]{revtex4}
\usepackage{color}
\usepackage{graphics}
\begin{document}

\title{Inversion of the Diffraction Pattern from an Inhomogeneously Strained Crystal\\ using an Iterative Algorithm}

\author{A.~A.~Minkevich}

\email[]{Andrey.Minkevich@univ-cezanne.fr}

\author{M.~Gailhanou}

\affiliation{TECSEN, UMR CNRS 6122 Universit\'{e} Paul
C\'{e}zanne, 13397 Marseille Cedex 20, France}

\author{J.-S.~Micha}

\affiliation{UMR SPrAM 5819, CEA-D\'epartement de Recherche
Fondamentale sur la Mati\`ere Condens\'ee, F-38054 Grenoble Cedex
9, France}

\author{B.~Charlet}

\affiliation{LETI, CEA Grenoble, 38054 Grenoble Cedex 9, France}

\author{V.~Chamard}

\author{O.~Thomas}

\affiliation{TECSEN, UMR CNRS 6122 Universit\'{e} Paul
C\'{e}zanne, 13397 Marseille Cedex 20, France}

\date{\today}

\begin{abstract}
The displacement field in highly non uniformly strained crystals
is obtained by addition of constraints to an iterative phase
retrieval algorithm. These constraints include direct space
density uniformity and also constraints to the sign and
derivatives of the different components of the displacement field.
This algorithm is applied to an experimental reciprocal space map
measured using high resolution X-ray diffraction from an array of
silicon lines and the obtained component of the displacement field
is in very good agreement with the one calculated using a finite
element model.

\end{abstract}

%PACS: 61.10.Nz; 62.20.-x; 42.30.Rx.\\

\maketitle
\section{Introduction}

The need to understand the physical properties of micro- and
nano-crystals leads to a fast development of techniques aimed at
probing the local structure. In addition, small objects have much
higher yield stresses as compared to their bulk counterparts
\cite{Arzt01} and the vicinity of surfaces and interfaces implies
strongly inhomogeneous stress fields. The experimental
determination of the local strain remains, however, an open issue:
electron microscopy has the required spatial resolution but
suffers from the need for sample thinning down to electron
transparency, which modifies the strain field \cite{Treacy85};
High Resolution X-Ray Diffraction (HRXRD) is both strain sensitive
and non-destructive but, as the phase of the scattered field is
not experimentally accessible, the strain profile at the nanometer
scale is only achieved through a model dependent approach
\cite{Baumbach99, Shen97}. In this context, direct inversion based
on x-ray diffraction is a rapidly progressing technique
\cite{Miao99, Pfeifer06, Nikulin98}. The possibility to directly
determine the structure from a diffraction pattern alone was first
mentioned by Sayre \cite{Sayre52}. It is based on the
"oversampling" conception, which allows to recover all Fourier
components of an object as soon as the diffracted intensity
pattern is sampled with a rate $\sigma$ at least twice the highest
frequency, namely the Nyquist frequency, which corresponds to the
object size. The direct space electron density is retrieved with
an iterative algorithm, first proposed by Gerchberg and Saxton
\cite{Gerchberg72} and further developed by Fienup
\cite{Fienup82}, relying on the fact that missing phases in most
cases for two or more dimensional data can be uniquely recovered
from the oversampled intensity information \cite{Bates82}. Since
then many versions of this algorithm were proposed as well as the
comparisons of their convergence behaviour properties
\cite{Elser03, Bauschke02, Marchesini07}. The method is based on
back and forth fast Fourier Transforms (FT) together with a set of
constraints in both direct and reciprocal spaces. As it is based
on FT, it is valid when the kinematical scattering at far field
takes place. This approach has been very successful in yielding
the density distribution of noncrystalline materials \cite{Miao99}
and crystals \cite{Williams06}. The strain distribution is more
difficult to retrieve since an effective complex-valued density is
used, where the amplitude is the density of the unstrained crystal
and the phase is approximately given by the scalar product of the
displacement $\vec{u}$ with the reciprocal Bragg vector
$\vec{G}_{hkl}$ \cite{Takagi69}. In this case, the convergence of
the existing algorithms is often problematic and has hindered so
far the general applicability of inversion to the diffraction of
strained objects. For some special shapes, the convergence may be
achieved \cite{Fienup87}. The first success concerning the case of
a weakly strained nano-crystal has been recently obtained
\cite{Pfeifer06}, but direct inversion of a diffraction pattern
from a very non uniformly strained crystal remains an unsettled
problem.

In this article, we present a new method based on modifications of
standard iterative algorithms, where additional constraints on the
spatial phase variations and on the crystal density uniformity are
introduced. The algorithm is successfully tested on experimental
data, where the displacement field retrieved from the x-ray
diffraction pattern measured on a highly non uniformly strained
crystal is in excellent agreement with the one calculated by
finite element modeling based on continuum elasticity. The details
of the developed iterative algorithm are explained in Section II.
Section III contains the sample description and high-resolution
diffraction experiment. Finally, the result of the inversion
obtained with our algorithm is presented and discussed in the
Section IV.

\section{Method for strained crystal density imaging}

The Error Reduction (ER) \cite{Fienup82} and the Hybrid
Input-Output (HIO) \cite{Fienup82, Millane97} iterative algorithms
are standard inversion techniques used in so called lens-less
x-ray microscopy. They are iteratively and cyclically used
together with a set of direct and reciprocal spaces constraints.
At each algorithm iteration $k$ the difference between the
calculated intensities and the experimental ones is expressed as:

\begin{eqnarray}
\label{3} E^2_k = \frac{\sum^{N}_{i=1} ( |F^{calc}_i| -
\sqrt{I^{meas}_i} )^2}{\sum^{N}_{i=1} I^{meas}_i},
\end{eqnarray}

\noindent where $|F^{calc}_i|$ is the magnitude of the calculated
amplitude and $I^{meas}_i$ is the measured intensity of point $i$
in the Reciprocal Space Map (RSM).

Unfortunately in general the application of this method is limited
to reconstructing a real positive valued function \cite{Fienup90}.
When the crystal is strained its direct space density is not real
and the positivity constraint cannot be applied. For a direct
space complex-valued density, only the finite size support
constraint defining the object size may be used. Very often this
constraint alone cannot provide reliable convergence of the
method. The stagnation in iterative scheme appears without finding
the correct solution. Such situation was observed for our case of
inhomogeneously strained Silicon on Insulator line (see Section
IV). Therefore without additional \emph{a priori} knowledge the
"phase problem" proved to be difficult to solve by this approach
in the case of inhomogeneously strained crystals. In this context,
the introduction of some additional constraints to the direct
space is mandatory. The cost of these constraints has, however, to
remain small \emph{i.e.} they should not require a fine
pre-knowledge of the crystal to reconstruct. The approach
presented in this article is restricted to the case of 2D plane
strain systems and to chemically homogeneous crystals, although a
generalization to 3D systems might be foreseen. In order to lift
the convergence problems, the following additional constraints in
direct space were added:

%-----------------Fig. 1---------------------------------
\begin{figure}
\includegraphics{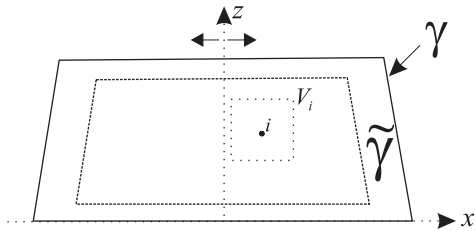}
\caption{\label{figLine} Schematic sketch of the crystal cross
section. $V_i$ is the area around point $i$, $\tilde{\gamma}$
corresponds to a small region near support ($\gamma$) edges.}
\end{figure}

I. The density, defined by direct space amplitudes, must be
uniform inside the support $\gamma$, except for a small region
near the support edges $\tilde{\gamma}$, where it decreases in the
external direction from the support (Fig.~\ref{figLine}):

\begin{multline}
\label{4} a_{k+1}(i) = \\
 \left \{ \begin{array}{ll}
a'_{k}(i), & |a'_{k}(i)-c_k(i)|<\epsilon  \\
a_{k}(i)+\beta (c_k(i)-a'_{k}(i)), & |a'_{k}(i)-c_k(i)|>\epsilon
%\\= \left \{\begin{array}{ll}a'_{k}(i) & a'_{k}(i) < a_{k+1}(i^{prev}) \\
%a_{k+1}(i^{prev}) & a'_{k}(i) > a_{k+1}(i^{prev})
%\end{array}\right. & i \in \tilde{\gamma}
\end{array} \right., \\
 c_k(i)=\left \{ \begin{array}{lll} 0, & i \notin \gamma  \\
\frac{\sum_{\tilde{i} \in V_i} a_k(\tilde{i})}{N_{V_i}}, & i \in
\gamma \setminus \tilde{\gamma} \\
a'_{k}(i), & i \in \tilde{\gamma}
\end{array} \right.,
\end{multline}

\noindent with $a_k(i)$ is the amplitude of point $i$ of input at
the $k$ iteration, $a'_k(i)$ is the amplitude of point $i$ of
output at the $k$ iteration, the input
$g_{k+1}(i)=a_{k+1}(i)e^{i\phi_{k+1}(i)}$ at the next iteration is
taken from the output of the previous one by implementation of
direct space constraints (for definitions of input and output see
Ref.~\onlinecite{Fienup82}), $\beta$ is the parameter, which is
taken in the [0.5, 1.0] interval, $\epsilon$ is the parameter
defining the threshold for applying the constraints (\ref{4}) to
each individual point $i$ in the Direct Space Map (DSM), $V_i$ is
the vicinity of point $i$, defining the set of neighbouring points
around point $i$, $N_{V_i}$ is the number of points in the $V_i$,
$\tilde{\gamma}$ is the narrow edge of support, the depth of this
edge is defined by fittable parameters. The shape of $V_i$ is not
very important (disc, square, etc.). The number of points inside
$V_i$ should not, however, be too small. The amplitude profile
inside the edge $\tilde{\gamma}$ is constructed automatically by
the iterative algorithm.

It was found that the constraint to amplitudes uniformity only is
insufficient to provide convergence for the data presented in this
article. This constraint works (it was checked on modeled
simulated data) only when the strain field inhomogeneity defined
by the variation range of displacement field derivatives is small
enough. When the amplitude of the derivatives variation increases,
this constraint alone does not allow to find the solution anymore.
For this reason a second constraint was added.

II. This second constraint is related to the maximum value that
the components of the discrete displacement derivatives
$\frac{\Delta_p u_j}{\Delta x_p}$ can take ($\Delta x_p$ is a step
along the $p$ direction of the DSM defining the spatial resolution
in this direction). This maximum value limits the possible phase
difference between neighbouring points in the DSM:

\begin{eqnarray}
\label{5} \left | \phi_{k+1}(i) - \phi_{k+1}(i') \right | <
G_{hkl} \Delta_p u_j^{max},
\end{eqnarray}

\noindent where $\Delta_p u_j^{max}$ is a maximum difference in
displacement component $u_j$ between neighbouring points $i$ and
$i'$ along $p$ direction.

To make this constraint more efficient, it is necessary to define
a minimum distance over which the displacement derivatives
$\frac{\Delta_p u_j}{\Delta x_p}$ sign is constant. These
distances depend on the particular properties of the sample, such
as shape, symmetry, origin of strain, etc. and they can be fitted
during an iterative process.

To implement these constraints in the iterative algorithm the
maximum value of the displacement derivatives, namely the
magnitudes $\Delta_p u_j^{max}$ have to be estimated. It was found
that for the iterative algorithm a precise knowledge of the value
of the maximum derivative is not very important. If the condition
(\ref{5}) is satisfied for most of the object volume, the
constraint to the phases (\ref{5}) should be switched off at the
last cycles of the iterative algorithm. Alternatively instead of
an estimation of the value of $\Delta_p u_j^{max}$, a
trial-and-error procedure can be easily performed.

\section{Experiment}

\begin{figure}                %Fig. 2
\includegraphics{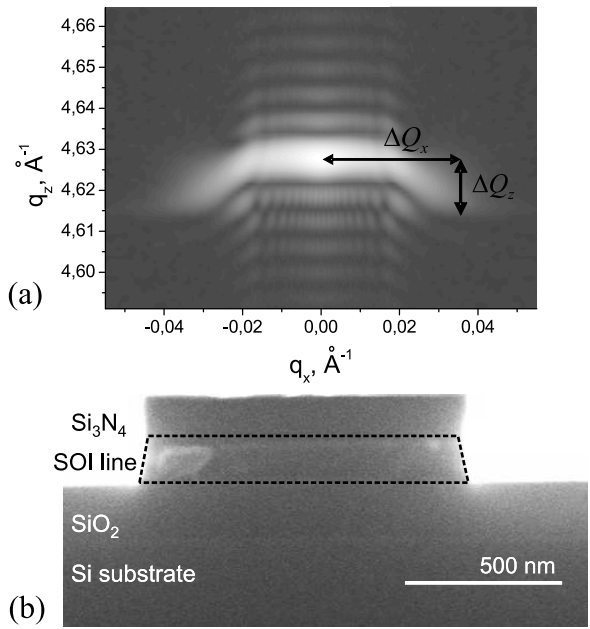}%
\caption{\label{figRSM+SEM} (a) RSM measured near 004 reflection
from Si lines on SiO$_2$/Si substrate. The intensity scale is
logarithmic. (b) Scanning electron microscopy image of the cross
section of the corresponding sample. The investigated Si on
Insulator (SOI) line is emphasized by a dotted line contour for
sake of clarity.}
\end{figure}

Here we describe high resolution diffraction measurements in the
vicinity of 004 Bragg reflection from Si lines on SiO$_2$/Si
substrate \cite{Gailhanou07}. The Si lines were obtained by
etching through a 160 nm thick Si$_3$N$_4$ mask a blanket 100 nm
SOI (silicon on insulator) film which lies on the top of a 200 nm
BOX (Buried Oxide) layer (Fig.~\ref{figRSM+SEM}b). The direction
of the lines is parallel to the [010] direction of the SOI crystal
and to the [110] direction of the substrate crystal. The system is
chemically homogeneous and uniform in the direction of the lines
and the strain field is therefore essentially 2D. The 2D high
resolution diffraction pattern was measured on the french CRG
beamline BM32 at ESRF around the 004 Bragg reflection
(Fig.~\ref{figRSM+SEM}a) in the plane $(x, z)$ normal to the
sample surface and to the SOI lines ($x
\parallel q_x
\parallel$ SOI [100] direction, $z \parallel q_z \parallel$ SOI
[001] direction, $y \parallel$ SOI line). Here $q_x$ and $q_z$ are
the components of $\vec{q} = \vec{k}_f - \vec{k}_i, |\vec{k}_f| =
|\vec{k}_i| = \frac{2\pi}{\lambda}$, $\vec{k}_i$ and $\vec{k}_f$
are the incident and diffracted wavevectors respectively. The
wavelength $\lambda = 1.54$\AA \, was selected using a Si (111)
monochromator. The SOI sample was mounted at the center of a (2+2)
circle diffractometer with a triple-bounce Si (111) analyzer and a
NaI:Tl scintillator. The beam size at the sample position was
500$\times$500 $\mu$m$^2$. The [001] directions of both crystals
are misaligned by 0.4$^{\circ}$. This small off-orientation
enables the measurement of the diffraction pattern of SOI lines in
the vicinity of $\vec{G}_{004}$ Bragg vector without any overlap
from the intensity scattered from the substrate. It is important
to note on Fig.~\ref{figRSM+SEM}a the absence of periodic
truncation rods expected from the interferences between lines in
spite of an X-ray coherence in this direction length of around
6~$\mu$m, which is much larger than the 2~$\mu$m line period. This
is due to important random phase shifts between lines, which might
be related to very small (1\AA~range) - but comparable to the
inverse of the scattering vector - ripples at the SiO$_2$ surface
\cite{Hartwig02}. Since a lot of perfectly uniform lines are
illuminated by the incident beam with a coherence length much
larger than the line width, the measured 2D diffraction pattern is
equivalent to the diffraction pattern of a single SOI line. In the
case of the 004 Bragg reflection the measured diffraction pattern
contains only the information on the $u_z$ component of the
displacement field $\vec{u}$. In order to be able to apply the
iterative inversion algorithm to the measured intensity it is
necessary to satisfy the oversampling condition $\sigma_{x,z} =
\frac{M_{x,z}}{L_{x,z}} > 2$ in the DSM by choosing the
appropriate step $\Delta q_{x,z} = \frac{2 \pi}{M_{x,z}}$ in the
RSM, where $M_{x,z}$ is the size of the DSM and $L_{x,z}$ is the
expected size of the object (support) in the corresponding
dimension $x$ or $z$. It has been shown \cite{Miao98} that it is
unnecessary to have an oversampling ratio $\sigma > 2$ in each
dimension to retrieve 2D and 3D objects. However, practically for
more reliable reconstructions it is better to have both
$\sigma_{x,z} >> 2$. In our experiment the oversampling ratio was
chosen to be $\sigma_z \approx 7.8$ and is related to the number
of measured points per thickness oscillation along $q_z$ in the
RSM. In the $x$ direction the experimentally chosen step $\Delta
q_x$ corresponds to an oversampling ratio $\sigma_x \approx 6.3$.
The steps $\Delta x \approx 7.9 \text{ nm}, \Delta z \approx 8.4
\text{ nm}$ in the DSM define the attainable resolution and relate
to the size of the RSM. The size of the RSM is restricted by the
area where the signal is above the background.

%-----------------Fig. 2---------------------------------

\section{Strain field reconstruction: results and analysis}

\begin{figure}                %Fig. 3
\resizebox{8.9cm}{7.3cm}{\includegraphics{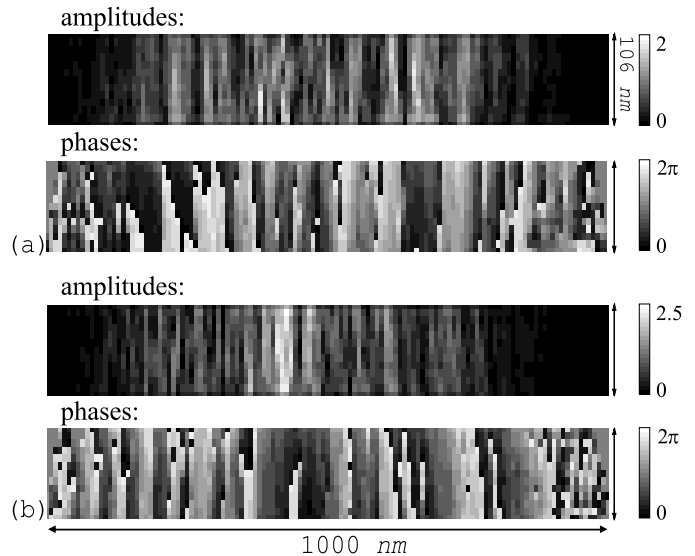}}
\caption{\label{fig3} (a) and (b) - two typical solutions of the
inverse problem using standard algorithms (ER + HIO): amplitudes
(in arbitrary units), corresponding to the shape and density of
SOI line, and phases (in radians).}
\end{figure}

The data in Fig.~\ref{figRSM+SEM}a were analyzed using the
iterative phase retrieval algorithm. Each time the algorithm was
applied with a new set of random phases in RSM. Two typical
results from the standard algorithm with support constraint, which
exactly correspond to the shape of SOI line, are shown on the
Fig.~\ref{fig3}. Both amplitudes and phases represent modulated
profiles with random behaviour, which do not have any physical
meaning. Moreover, the results are not consistent from one pass of
algorithm to another. However, the error metrics are very small
and vary in the interval $E^2_k \sim [1.7 * 10^{-3}, 2
* 10^{-3}]$. It means that, in the case of an inhomogeneously
strained crystal expressed as a complex-valued density, different
combinations of amplitudes and phases in direct space can yield
very similar FT amplitudes images. They correspond to local minima
with very small error metric (\ref{3}) causing the appearance of
an ambiguity in the solutions. This is especially the case for
experimental data where, because of the presence of noise, the
difference between error metrics of correct and local minima
solutions vanishes. Once the algorithm reached the local minima
with such a small error metric, it stagnates near it, \emph{i.e.}
further iterations do not produce any significant changes. In this
case the zero density region outside the support, arising from
signal oversampling, cannot compensate for the unknown phases in
the diffraction pattern, because the values in this area become
negligibly small.

Experimental artifacts are not a reason for the non-convergence as
a study was also done on noise-free modeled simulated data leading
to the same conclusion. This is a mathematical problem of
convergence of iterative algorithms in the case of complex-valued
objects. It is also interesting to point out that when the values
of deformations were artificially reduced to the values similar to
those reconstructed in Ref. \onlinecite{Pfeifer06}, the solution
was found by standard methods. This clearly shows that standard
iterative algorithms are not effective for the case of highly
inhomogeneously strained crystals.

We now describe the application of the previously described
modified algorithm to the same experimental data. The direct space
constraints, discussed in the first part of this article, were
added to the iterative phase retrieval algorithm in addition to
the standard support constraint. Supplementally to these direct
space constraints, the symmetry property of the displacement field
with respect to the vertical axis $z$ shown on Fig.~\ref{figLine}
was used. An estimation of the maximum values of the displacement
derivatives $\frac{\Delta_x u_z^{max}}{\Delta x}$, $\frac{\Delta_z
u_z^{max}}{\Delta z}$ ($\Delta x$, $\Delta z$ are steps along $x$
and $z$ of the DSM respectively) was also carried out from
experimental data. The corresponding magnitude of the derivative
of $u_z$ along $x$ is found from the distance $\Delta Q_x$, shown
on Fig.~\ref{figRSM+SEM}a, via the relation $\Delta Q_x = G_{004}
\frac{\Delta_x u_z^{max}}{\Delta x}$. This value corresponds to a
$\frac{3}{4}\pi$ maximum phase difference in the DSM between
neighbouring points along $x$. Similarly the magnitude for the
maximum displacement derivative $u_z$ along $z$, $\frac{\Delta_z
u_z^{max}}{\Delta z}$, is calculated from the distance $\Delta
Q_z$, shown on Fig.~\ref{figRSM+SEM}a, via the relation $\Delta
Q_z = G_{004} \frac{\Delta_z u_z^{max}}{\Delta z}$. The validity
of these estimations was checked at the end of the inversion.

%-----------------Fig. 3---------------------------------

\begin{figure}               %Fig. 4
\resizebox{8.9cm}{5.7cm}{\includegraphics{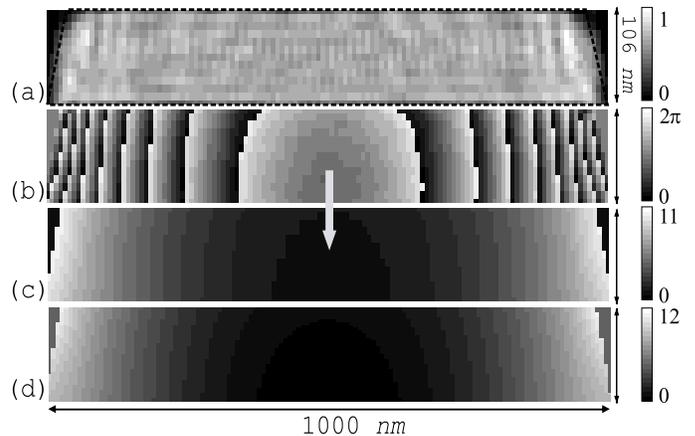}}
\caption{\label{fig4} The solution of the inverse problem: (a)
amplitudes (in arbitrary units), which represent shape and density
of SOI line: the dotted line trapezoid corresponds to the
trapezoid of Fig.~\ref{figRSM+SEM}b, (b) phases (in radians), (c)
retrieved displacement field $u_z$ (in \AA), (d) displacement
field $u_z$ calculated by finite element modelling (in \AA)
\cite{Gailhanou07}.}
\end{figure}

A good starting point for the shape was obtained from measurements
on the scanning electron micrograph. The thickness oscillations
along the $z$ direction, which are clearly observed in the RSM,
because of the small strain gradient in this direction, provided
also a good estimation. The adaptive shrinkwrap procedure
\cite{Marchesini03}, which starts from a larger support and
gradually removes from it the pixels whose amplitudes tend towards
zero, does not work in our case. For this reason a support fitting
procedure was developed, in order to gradually change the support
area $\gamma$ and the edge of the support $\tilde{\gamma}$ during
the iterations. Generally one cycle of iterative algorithm
included consecutively 50 iterations of ER, 50 iterations of HIO,
50 iterations of modified HIO (\ref{4}) with phase constraints
(\ref{5}) and 50 iterations of modified HIO (only constraints to
amplitudes (\ref{4})). If the support is very well defined
(\emph{i.e.} it corresponds to the exact shape of the object), the
finding of the solution takes about 4-8 cycles. Many trials of the
algorithm were performed starting from random phases in RSM and
each time converged to the solution with $E^2_k \approx 10^{-3}$.
With respect to the accuracy of the data all the solutions are the
same, and one of them is plotted in Fig.~\ref{fig4}. The
discontinuities in the retrieved phases map are related to the
crossing of phases through $2\pi$ ($2\pi$ corresponds to a
displacement equal to $\frac{2\pi}{G_{004}}$). Some strain
$\epsilon_{zz}$ along $z$ is also found gradually appearing at the
edges of the line, in the region where the derivative
$\frac{\partial u_z}{\partial x}$ becomes very high. This strain
causes the appearance of the "moustache" shape intensity
distribution in the RSM. There is also a small amount of
homogeneous strain in the $z$ direction, which was found from the
difference in $q$ between the Bragg maxima of SOI lines and Si
substrate. Using the retrieved phases the displacement field was
calculated (Fig.~\ref{fig4}c). The maximum value of the
displacement $u_z$ is about 11\AA.

The main advantage of this approach is its model independence as
opposed to the case of, for example, finite element calculations.
Such calculations were also done for this sample, considering the
residual stress in the Si$_3$N$_4$ top layer as the reason for the
strain appearance in the line \cite{Gailhanou07}. The displacement
fields obtained by these two approaches are in very good agreement
(Fig.~\ref{fig4}c,d), which definitely validates our inversion
procedure.

\section{Conclusion}

We have presented a modified iterative algorithm with additional
direct space constraints. It is shown that the displacement field
in a highly inhomogeneously strained crystal can be retrieved from
its x-ray diffraction pattern alone. This algorithm is applied to
retrieve the displacement field inside SOI lines from experimental
data. In this particular case, finite element modeling was also
possible and quite successful in describing the displacement
field, and a very good agreement between the two methods was
found. This opens important perspectives for local strain
determination at the nanoscale, in particular for the cases where
the model dependent approach cannot be used.

\begin{acknowledgments}
The author A.A.M. is very grateful to S.Labat for discussions. The
ESRF is acknowledged for beamtime allocation.
\end{acknowledgments}

%\bibliography{inversion}% Produces the bibliography via BibTeX.

\end{document}